\documentclass{PoS}
\usepackage{subfigure}
\title{Charm and strange hadron spectra from overlap fermions on HISQ gauge configurations}

\ShortTitle{Strange and charm hadron spectra from overlap fermion}

\author{S. Basak$^a$, S. Datta$^b$, Padmanath M.$^b$, P. Majumdar$^c$, and 
{\speaker{N. Mathur}}$^b$}

\author{{\hspace*{1.0in}}(Indian Lattice Gauge Theory Initiative)\\\\
\llap{$^a$}School of Physical Sciences, National Institute of Science Education 
and Research, Bhubaneswar 751 005, India\\
\llap{$^b$}Department of Theoretical Physics, Tata Institute of Fundamental Research, Homi Bhabha Road, Mumbai 400005, India\\
\llap{$^c$}Department of Theoretical Physics, Indian Association for the Cultivation of Science, Kolkata 700032, India.
\\
        E-mail: \email{nilmani@theory.tifr.res.in}}

\abstract{We report here results on charm and strange hadron spectra. 
  Adopting a mixed action approach, we use
  overlap fermions for valence quarks, on a background of 2+1+1
  flavours HISQ gauge configurations of MILC collaboration. Two lattice spacings (0.09 fm
  and 0.06 fm) are used. We find the hyperfine splitting of 1S
  charmonia to be $114^{+3}_{-2}$ MeV and $109^{+4}_{-3}$ MeV, and
  the splitting ($m_{\Omega_{ccc}}-{3\over2}m_{J\//\Psi}$) is found to
  be $110_{-10}^{+20}$ MeV and $120(10)$ MeV, corresponding to lattices
  with spacings $a$ = 0.09 and 0.06 fm respectively.  
  We also look at the ratio of the leptonic decay constant $f_{D^{*}_s}/f_{D_s}$.}

\FullConference{The 30 International Symposium on Lattice Field Theory - Lattice 2012,\\
		June 24-29, 2012\\
		Cairns, Australia}

\begin{document}
\vspace{-0.1in}
\section{Introduction}
During the last few years MILC lattice collaboration 
has generated~\cite{HISQ_MILC} a large set of configurations with the one-loop, 
tadpole improved Symanzik gauge action and the highly improved
staggered quark (HISQ) fermion action~\cite{HISQ_action}.  Four
flavours of dynamical sea quarks: degenerate up and down, strange, and
charm, were included in these configurations. Taste violations in the HISQ action were found to be small~\cite{HISQ_action}. On the
other hand, overlap fermion action~\cite{overlap} has exact chiral
symmetry~\cite{overlap,overlap_chsymmetry} on the lattice and is
automatically ${\cal{O}}(a)$ improved. It has also many desirable
features, such as the adaptation of multi mass
algorithms~\cite{overlap_multimass}.  However, using overlap action for the
dynamical quarks is still prohibitively costly, except with fixed
topology~\cite{overlap_dynamical}. In this work we have adopted a
mixed action formalism where we used overlap valence quarks on the
HISQ gauge configurations generated by MILC collaboration~\cite{HISQ_MILC}. 
By adopting such a mixed action approach, one can get advantage of the chiral symmetry and low
quark mass limit of overlap fermions, and the advantage of having a
large set of configurations with small discretization errors as well
as small taste breaking effects. One also gets the advantage of
simulating both light, strange as well as heavy fermions on the same
lattice formalism with chiral fermions having
no $\mathcal{O}(a)$ errors. Of course we will have to encounter the usual issues related to mixed action and partial quenching~\cite{MAPQchpt}. 
Similar approach has been taken by the $\chi QCD$
collaboration, who used overlap valence quarks on 2+1 flavours
dynamical domain wall gauge configurations~\cite{overlap_chiQCD1}. 
 
In this work, by using above mentioned mixed action approach, we
report our preliminary results on charm and strange hadron spectra as well as
leptonic decay constants for $D_s$ and $D^*_s$ mesons.

\vspace{-0.1in}
\section{Numerical details}
We used two sets of dynamical 2+1+1 flavours HISQ lattice ensembles,
generated by the MILC collaboration : a set of $32^3 \times 96$ 
lattices with lattice spacing $a = $ 0.09 fm, and a set
of $48^3 \times 144$ lattices with $a = $ 0.06
fm. The details of these configurations are summarized in Ref.~\cite{HISQ_MILC}. The results reported here were obtained from 100
configurations on the coarser lattice, and 35 configurations on the
finer lattice. Here we did not address the uncertainties in the determination 
of lattice spacings which we will address in subsequent works.

For valence quarks we used overlap action~\cite{overlap}. For the
numerical implementation of massive overlap fermions the methods used
by the $\chi QCD$ collaboration ~\cite{overlap_chiQCD} were followed.
To evaluate the sign function, the Zolotarev approximation is used,
and the low Wilson eigenmodes are projected out by Arnoldi method.  We
used 350 and 160  Wilson eigenmodes for the coarser and the finer lattices 
respectively. The constraint here came from the size of the memory of the machine.  Using multimass CG algorithm we were able
to input a range of quark masses covering from light, strange to heavy
quarks. It is worthwhile to mention that one should not use
the same precision for the evaluation of light and heavy quark
propagators. While light quark and strange quark propagators needs
relatively lower precision, for heavy quark propagators which decay
rapidly, having higher precision in the inversion is absolutely
necessary. Otherwise, one will get flat hadron correlators at the
middle of the lattice.  However, for using different precisions in the
multimass we found that it is not necessary to reload the eigenmodes
repeatedly, which increases the total inversion time. We used periodic
boundary condition in the spatial and antiperiodic in the temporal
directions. Gauge configurations were first fixed to coulomb gauge and
then HYP smeared (one HYP). Using both point and wall sources we
calculated various point-point, wall-point as well as wall-wall
correlators.
  
The strange mass was tuned by setting the $\bar{s}s$ pseudoscalar mass
to 685 MeV \cite{strange_tune}.  With the strange mass tuned this way,
the mass of the vector $\phi(\bar{s}s)$ meson was found to be 1050(20)
MeV and 1030(10) MeV on the coarser and finer lattices respectively,
and the mass of the $\Omega(sss)$ baryon was found to be 1620(40) and
1630(30) MeV respectively. The charm mass is tuned by setting the
spin-averaged 1S state mass, $(m_{\eta_c}+ 3 m_{J/\psi})/4$, to its
physical value.  The charm quark mass ($ma_{charm}$) in these lattices
are obtained to be 0.55 and 0.336, respectively. Interestingly, the
$\overline{MS}$ mass of charm, 1.275 GeV, corresponds to $ma$ = 0.582
and 0.388, respectively, indicating that the mass renormalization
constant $Z^{\overline{MS}}_m$ is close to 1 for this action.

Since $ma$ is not very small, we need to be careful about
discretization errors. Overlap action does not have $O(ma)$ errors. In
order to estimate the size of discretization errors coming from higher
orders of $ma$, we look at the energy-momentum dispersion relation of
the 1S charmonia. We calculated the mesons at various external momenta
$p^2 = (2 \pi/L)^2 n^2$, with $n \le 2$.  We used both point source
and a momentum-induced wall source for the finite momenta mesons. For
the wall source at finite momenta, we put a suitable phase factor in
the wall so as to project to the suitable momentum. While this
requires a separate inversion for each momenta, that is more than
compensated by the improvement in the signal, as shown in Fig. 1.  In both
  figures we show both point source and wall source results
  for the coarser lattice. Note that the wall source results are from
  8 configurations, but the statistical error is already comparable to
  that from 100 configurations with the point source.

In the left panel of Fig. 1, we show $E(p)$ for various momenta for 
the pseudoscalar meson on our coarser lattices. Also shown are the 
continuum dispersion relation, $E^2=m^2+p^2$, and the lattice 
dispersion relation for the standard scalar action with $O(ma)^2$ 
error, $\sinh^2(E(p)a/2) = (\sinh^2(ma/2))^2 + \sin^2(pa/2)$. While 
substantial deviation from the continuum dispersion relation is seen, 
the lattice scalar dispersion relation seems to explain the data quite 
well. For a more quantitative analysis, we follow the standard 
practice of introducing an effective ``speed-of-light'' $c$ through 
$E^2(p)=m^2+p^2 c^2$. The value of $c$ obtained using this relation 
and $E(p)$ at various $p$ are shown in the right panel of Fig. 1. As 
the figure shows, we get $c^2 \sim 0.75$ on our coarser lattices, and 
$c^2 \sim 0.89$ on the finer lattices. Using quenched overlap 
fermions, on an even coarser lattice, 
Ref. \cite{overlap_quenched_disp} found $c^2 \sim 1$ for the 
charm. However, on our dynamical lattices, we obtained $c^2$ values 
similar to that obtained in the literature with clover action, 
indicating a similar size of cutoff in the overlap action as that in 
the clover. The $c^2$ values are well-approximated by the estimate 
$ma/\sinh(ma)$, suggested by the lattice scalar action.

\begin{figure}[h]
\includegraphics[width=0.5\textwidth,height=0.33\textwidth,clip=true]{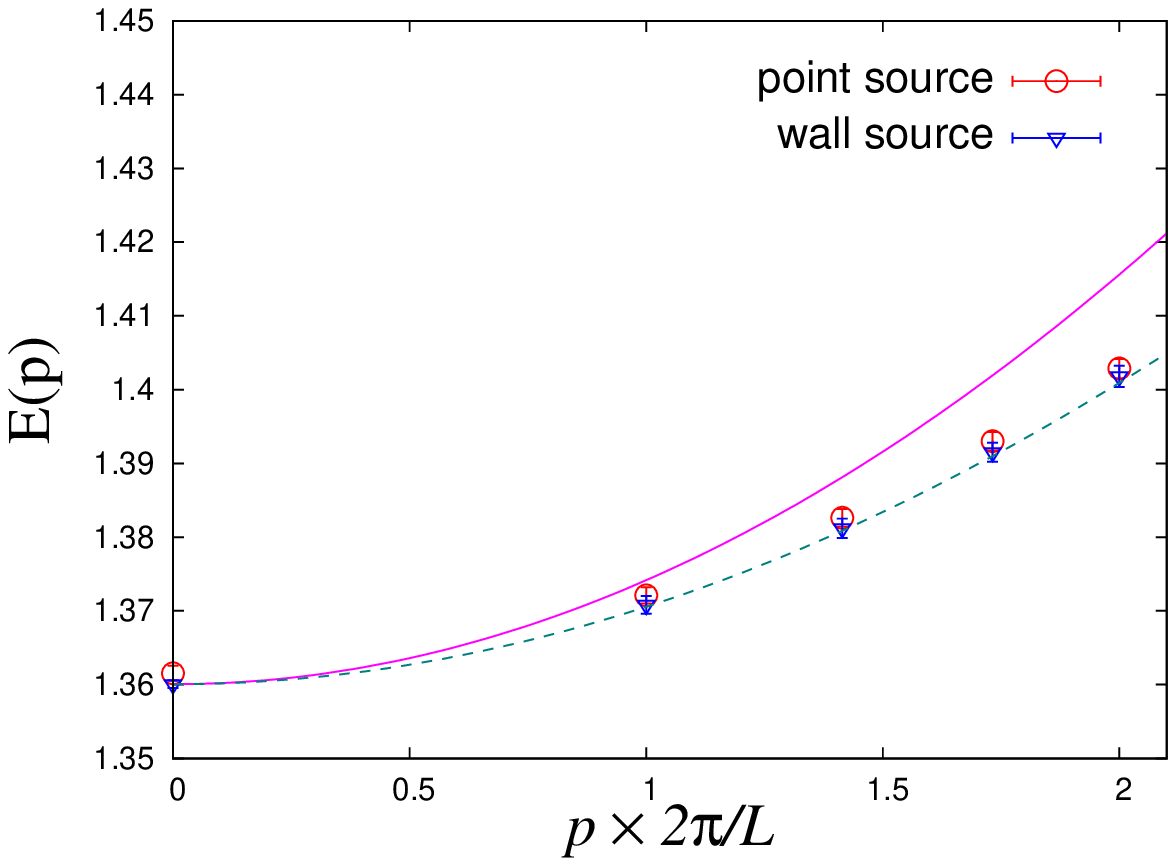}
\includegraphics[width=0.5\textwidth,height=0.33\textwidth,clip=true]{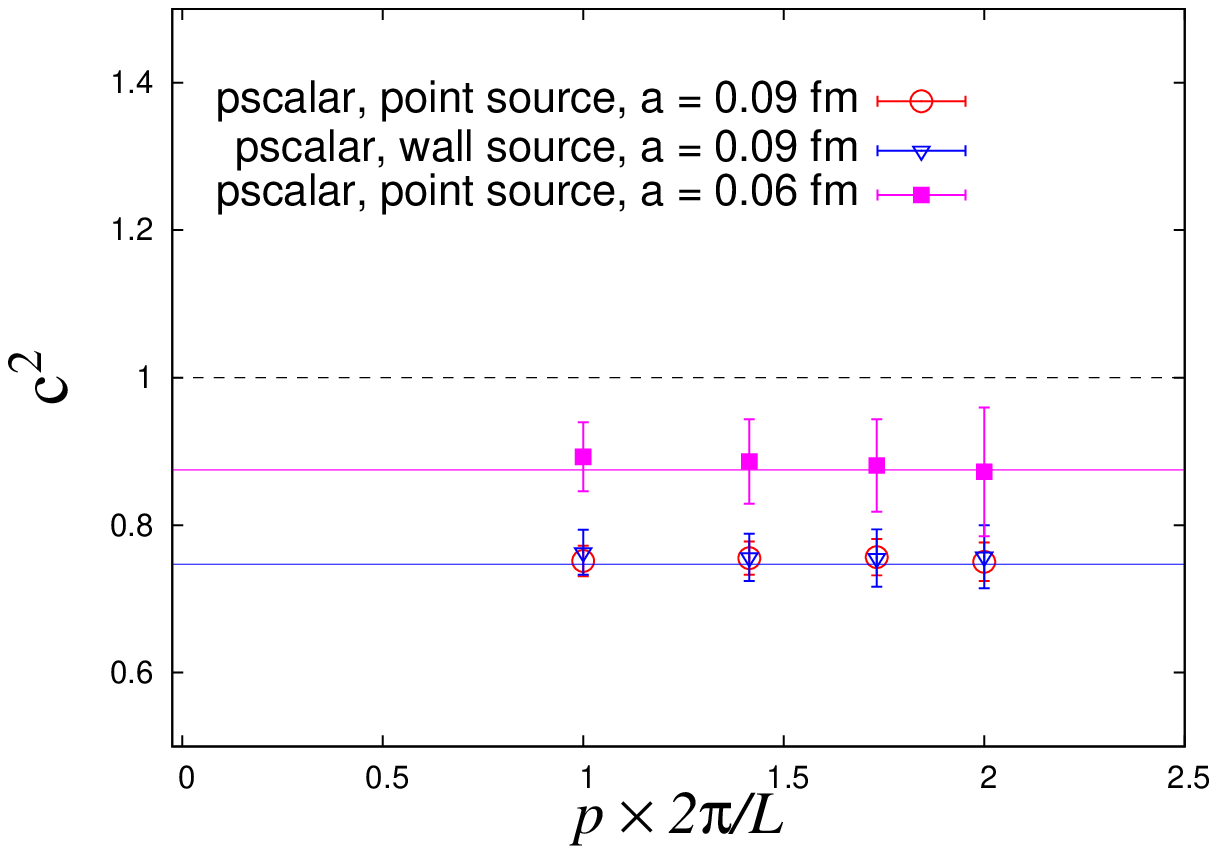}
\vspace{-0.17in}
\caption{(Left) Energy-momentum dispersion relation for the $\eta_c$
  on the lattices with $a$ = 0.09 fm. Also shown are the continuum
  relativistic dispersion relation (solid line), and the lattice dispersion
  relation (dashed line) for the standard scalar action. (Right)
  $c^2=(E^2(p)-E^2(p=0))/p^2$ calculated for $\eta_c$ at various
  values of $p$, on lattices with $a$ = 0.09 and 0.06 fm. 
  Solid horizontal lines represent the estimate $ma/\sinh(ma)$, suggested
by the lattice scalar action.}
\end{figure}

\vspace{-0.12in}
\section{Results}
We utilized both point and wall sources to generate a large set of
quark propagators.  Using those we calculated various hadron
correlators and extracted light, strange and charm hadron masses.  The
pseudoscalar meson masses ranged between $240-3582$ MeV for the
coarser lattices and between $420-5275$ MeV for the finer lattices.
In Fig. 2 we plot the pseudoscalar meson masses over a range of
quark masses. 

Using these correlators, we report here preliminary results for spectra
of charmonia, mesons of $D_s$ family, and charmed baryons. We also
report preliminary studies of leptonic decay constants of $D_s$
and $D^*_s$ mesons.
\begin{figure}[h]
\vspace{-0.1in}
\includegraphics[width=0.5\textwidth,height=0.28\textwidth,clip=true]{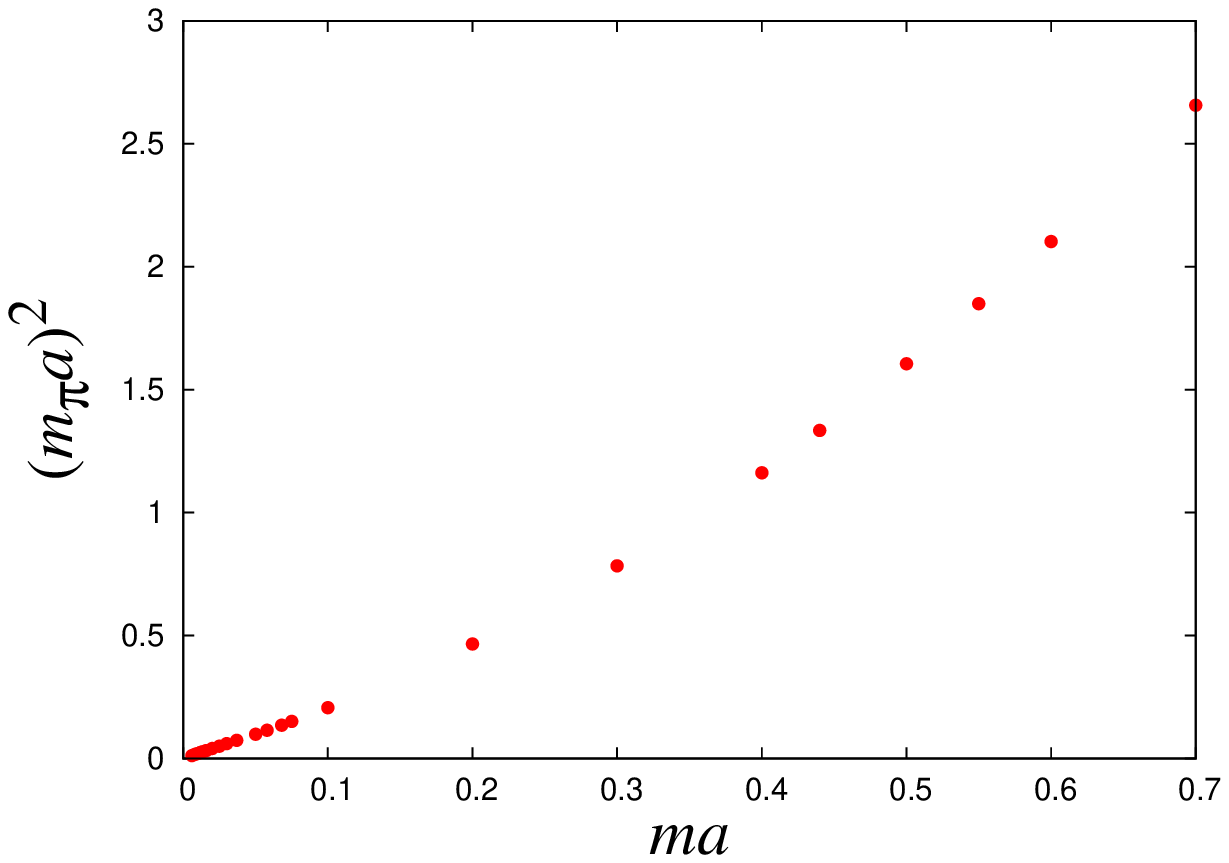}
\includegraphics[width=0.5\textwidth,height=0.28\textwidth,clip=true]{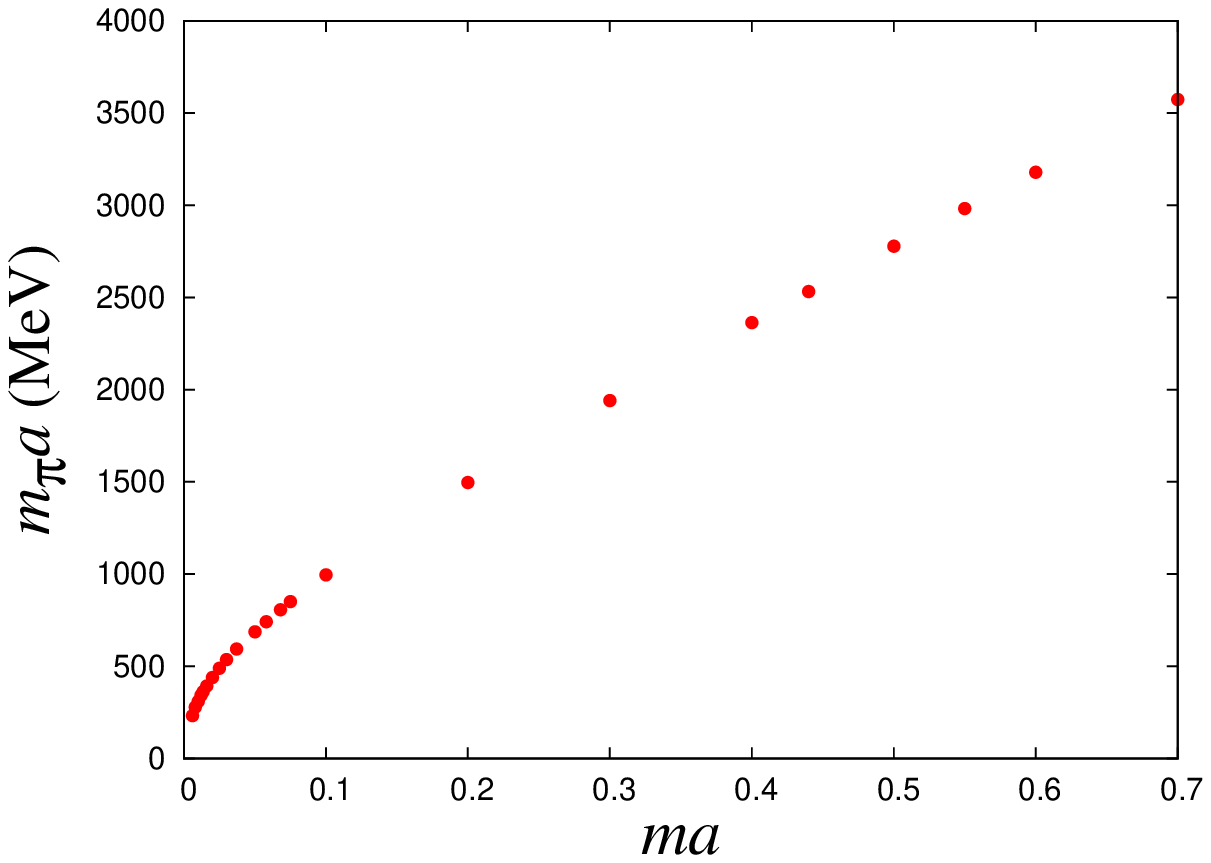}
\vspace{-0.17in}
\caption{Pseudoscalar meson masses over a range of light to heavy
  quark masses. Left figure is for $m_{\pi}^2a$ as function of quark
  mass $ma$, and the right figure is for $m_{\pi}a$ vs $ma$ in physical
  unit.}
\end{figure}

\vspace{-0.1in}

\subsection{Hyperfine splitting in 1S charmonia}
The hyperfine splitting in 1S charmonia is one of the most well
studied physical quantities in lattice charmonium calculations over
the year, and until very recently~\cite{MILC_hfs} lattice 
results were found to be smaller than the experimental value ($\sim$
116 MeV). This underestimation was mainly due to the discretization
error associated with the charm quark action and the quenched approximation. 
In our study we also
calculated this splitting. In Fig. 3, we showed the effective
splittings between vector and pseudoscalar correlators (jackknifed) at
the tuned charm mass for wall-point correlators. Horizontal lines shown are 
the fit results obtained by fitting {\it only} these correlators. 
Our final estimated results, by {\it simultaneously fitting} the wall-point
and point-point correlators, for this hyperfine splitting are
$114^{+3}_{-2}$ MeV and $109^{+4}_{-3}$ MeV corresponding to lattices
with spacings $a$ = 0.09 and 0.06 fm respectively. 
\begin{figure}[h]
\vspace{-0.15in}
\begin{center}
\includegraphics[width=0.47\textwidth,height=0.28\textwidth,clip=true]{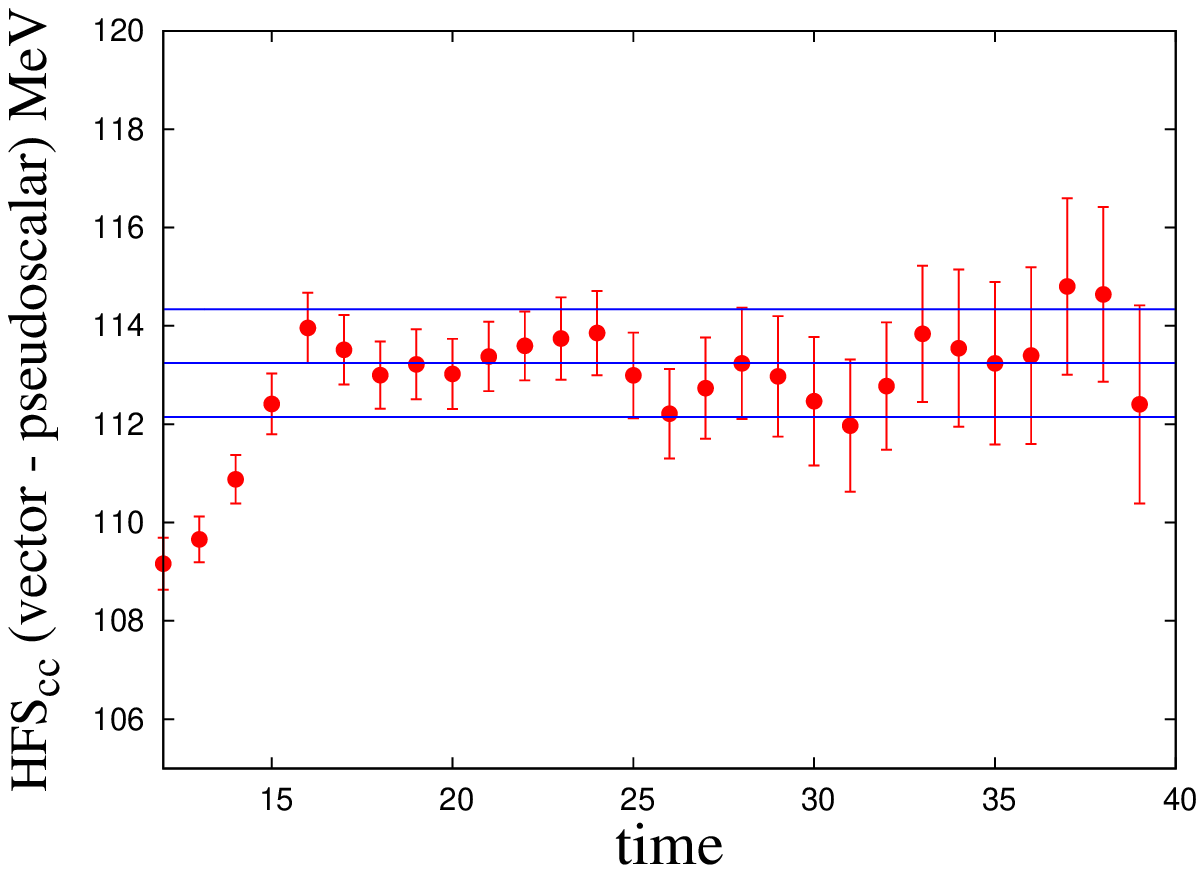}
\includegraphics[width=0.47\textwidth,height=0.28\textwidth,clip=true]{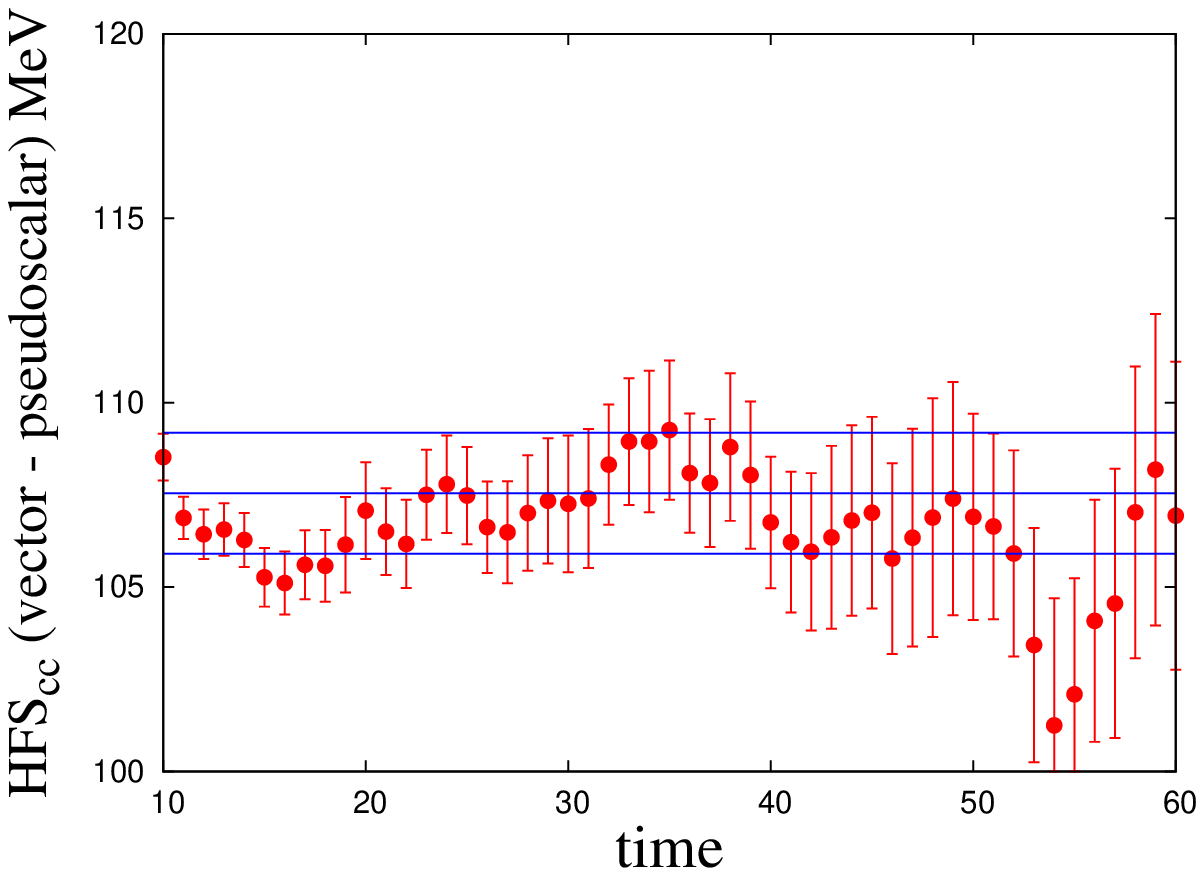}
\end{center}
\vspace{-0.2in}
\caption{Effective hyperfine splitting in 1S charmonia for wall-point correlators for lattices with spacing 0.09 (left) and 0.06 fm (right). Horizontal lines show the fit results obtained by fitting {\it only} these correlators.}
\end{figure}
\vspace*{-0.45in}
\subsection{Charm and strange mesons}
Along with pseudoscalar and vector mesons we also calculated the
ground state spectra of tensor, axial and scalar charmonia. 
Besides charmonia we also calculated the strange-charm
$D_s$ mesons. In Fig. 4(a) we showed splittings between these various
mesons at two lattice spacings. It 
is to be noted that the splitting of $D_s-\eta_c/2$, which has smaller 
lattice spacing error~\cite{davies_Ds_eta_c}, 
obtained is consistent with its experimental value.
\begin{figure}[h]
\vspace{-0.1in}
\subfigure[]{
\includegraphics[width=0.5\textwidth,height=0.35\textwidth,clip=true]{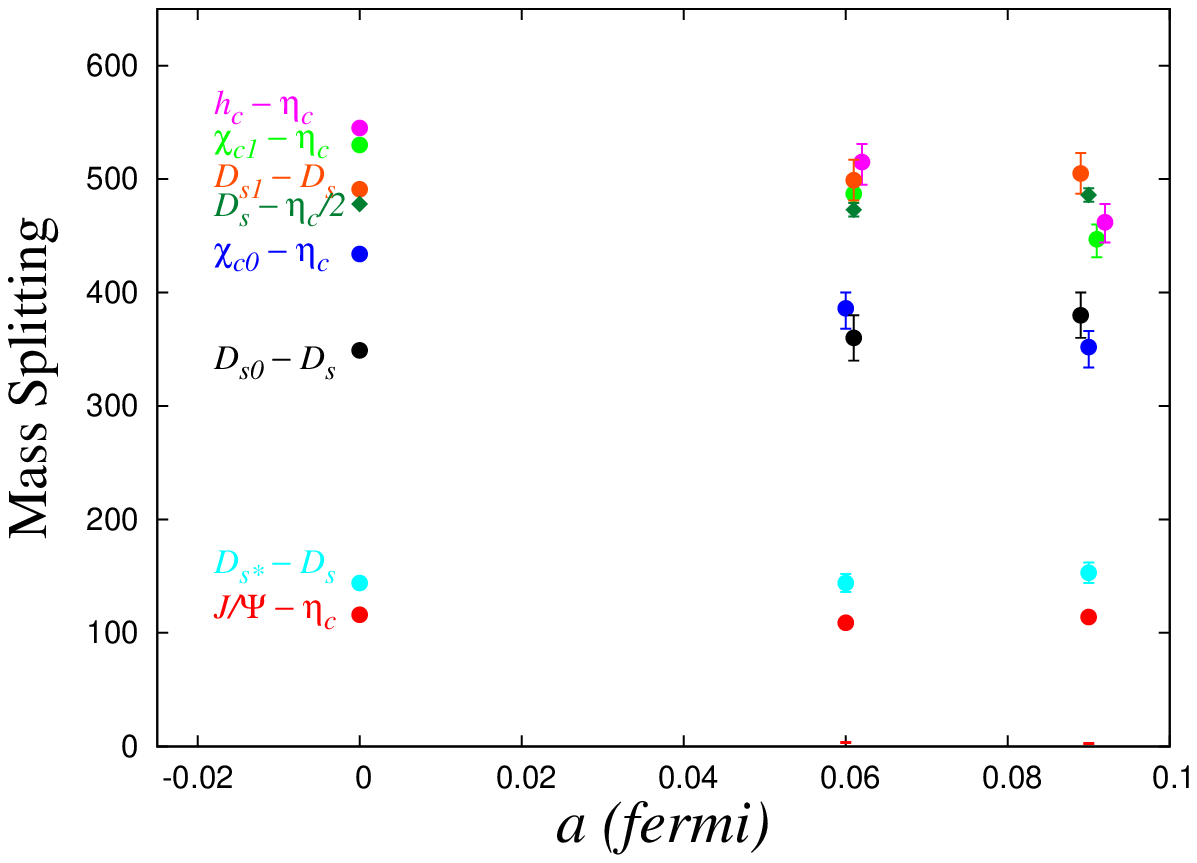}}
\subfigure[]{\includegraphics[width=0.5\textwidth,height=0.35\textwidth,clip=true]{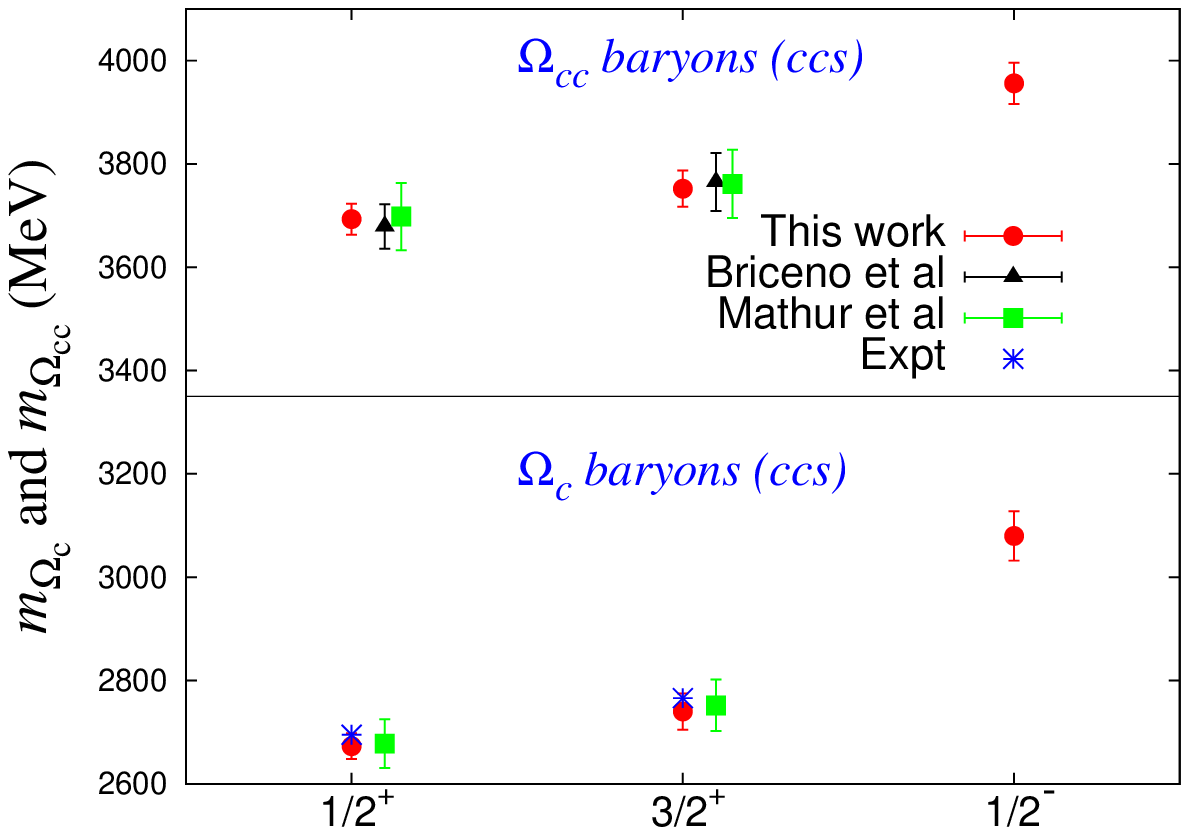}}
\vspace{-0.3in}
\caption{(a) Meson mass splitting for charmonia and strange-charm mesons at two lattice spacings. Experimental values are shown in the left side. (b) $\Omega_c$ and $\Omega_{cc}$ baryon masses. Also shown are other lattice determinations, and the experimental values, where available.}
\end{figure}
\vspace{-0.25in}
\subsection{Charmed baryons}
Recently, there have been exciting developments in heavy-baryon
physics, both theoretically and experimentally. Results from LHC and
future charm-bottom factories are expected to add to the excitement in
this field in the near future. On our lattices we extracted ground
state spectra of charmed baryons with one or more charm quark content,
for example, baryons with quark content $csu, cuu, css$ and $ccs$. In
fig. 4(b) we showed results for $\Omega_c(css)$ and $\Omega_{cc}(ccs)$
baryons.  It is to be noted that for these baryons we extracted masses
for both spin 1/2 and spin 3/2 with both parities, some of which are
yet to be measured experimentally. Our results are consistent with
other lattice results~\cite{charm_baryon_hwlin}, and the experimental
values, where available.  Results for other charm baryons at various
pion masses were shown in Fig. 5.  The chiral extrapolation will
involve effects from partial quenching as well as mixed action
formalism.  Mixed action partially quenched chiral perturbation theory
(MAPQ$\chi$PT) has been developed for various mixed action
formalisms~\cite{MAPQchpt} which requires calculation of
$\Delta_{mix}$ \cite{delta_mix}, the low energy constant representing
$\mathcal{O}(a^2)$ discretization dependence.  In future, utilizing
MAPQ$\chi$PT we will extrapolate these masses to the physical limit.
We also calculated the experimentally unknown triply charmed baryon
$\Omega_{ccc}$, and a better way to quote its mass is through the
splitting of $m_{\Omega_{ccc}}-{3\over2}m_{J\//\Psi}$, which we
extracted to be $110_{-10}^{+20}$ MeV and $120(10)$ MeV, for 0.09 fm
and 0.06 fm lattices respectively.
\begin{figure}[h]
\hspace{-0.2in}
\includegraphics[width=0.52\textwidth,height=0.36\textwidth,clip=true]{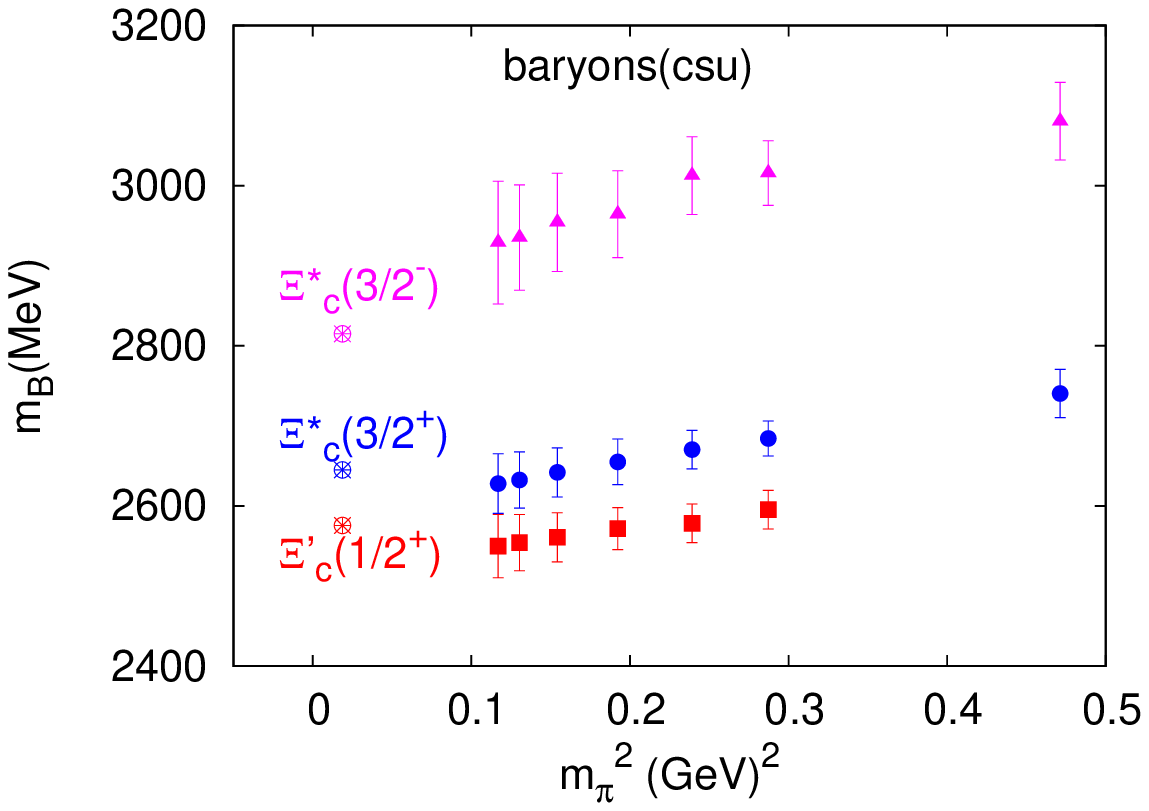}
\includegraphics[width=0.52\textwidth,height=0.36\textwidth,clip=true]{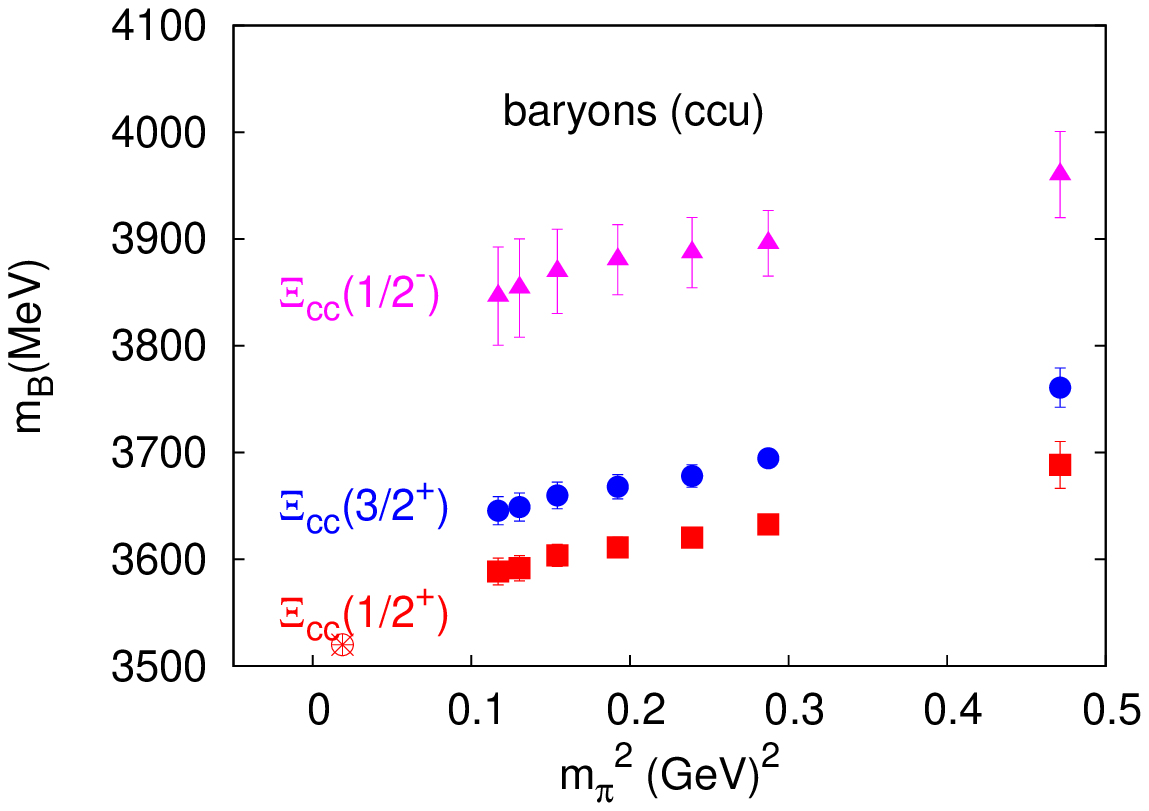}\\ 
\vspace{-0.4in}
\caption{Ground state spectra for charmed baryons at various pion masses for $32^3\times 96, a = 0.09 fm$  lattice.}
\end{figure}
\vspace*{-0.15in}
\subsection{Decay constants}
The decay constants of heavy-light mesons are experimentally very important:
they are essential ingredients in extracting CKM matrix elements from decays
of heavy-light mesons. For the suitable continuum currents
\( \{ A_\mu, \: V_\mu \} \ = \ \{ Z_A \: (\bar{s} \gamma_\mu \gamma_5 
c)_{\rm lat}, \ Z_V  \: (\bar{s} \gamma_\mu c)_{\rm lat} \}, \) 
where $Z_A, Z_V$ are the renormalization constants to match the lattice operators to continuum currents, the leptonic decay constants for the pseudoscalar 
$D_s$ and the vector $D_s^*$ are defined as 
\begin{equation}
\langle 0|A_{\mu}|D_S(p)\rangle = i \: f_{D_s} \: p_{\mu}, \qquad 
\langle 0|V_{\mu}|D^{*}_s(p,\lambda)\rangle = i \: f_{D^{*}_s} \: 
m_{D^{*}_s} \: \epsilon^{\lambda}_{\mu},
\end{equation} 
where $\epsilon_\mu^\lambda$ is the polarization vector for the vector
meson.  From the amplitude of the $\langle V(t) \: V(0) \rangle$
correlator, $ f_{D^{*}_s}$ can therefore be extracted using standard
techniques. Similarly, $f_{D_s}$ can be extracted from $\langle A_4(t)
\: P(0) \rangle$ and $\langle P(t) \: P(0) \rangle$ correlators, where
$P = Z_P \bar{s} \gamma_5 c$ is the pseudoscalar current.

Employing point-point as well as wall-point and wall-wall correlators
we extracted the suitable amplitudes for the above matrix elements. 
The relevant renormalization
constants, $Z_V$ and $Z_A$ have not been calculated yet. On the other
hand, since the overlap action possesses exact chiral symmetry on the
lattice, the ratio $Z_V/Z_A$ is expected to be close to 1 (it is
exactly 1 for massless quarks). This allows us to get an estimate of
the size of $f_{D^*_s}$ through the calculation of
\[ R(f_{D^*_s},f_{D_s}) \ = \ \frac{Z_A}{Z_V} \: \frac{f_{D^*_s}}{f_{D_s}} 
\ \approx \ \frac{f_{D^*_s}}{f_{D_s}}. \] 
since $f_{D_s}$ has been calculated
very precisely in the literature.  Our preliminary results for
$R(f_{D^*_s},f_{D_s}) = 1.15(10)$ on the lattices with $a = 0.09fm$.
 The mixed action effects will be smaller
for heavy-light mesons, and in the ratio its effects will be minimal. 
\vspace{-0.15in}
\section{Conclusions}
In this work we reported preliminary results on the ground state
hadron masses along with charm-strange meson decay constants by using
a mixed action approach, comprising overlap valence quarks generated
on the background of dynamical 2+1+1 flavours HISQ configurations.
The results, in particular the hyperfine splitting of 1S charmonia,
are encouraging and suggest that the overlap valence on 2+1+1
flavor HISQ configurations is a promising approach to do lattice QCD
simulation with light, strange and charm quark together in the same
lattice formulation. However, at these lattice spacings, 
the discretization errors of the overlap action for the charm, as 
evident from the dispersion relation, are not
negligible, and are similar in size to those of the clover action.

This is a continuing study and we expect to be able to do suitable
chiral and continuum extrapolations, to make experimentally relevant
predictions for various charmed
baryons. The splitting ($m_{\Omega_{ccc}}-{3\over2}m_{J\//\Psi}$),
between $J\//\Psi$ and the unknown triply charmed baryon
$\Omega_{ccc}$ was found to be $110_{-10}^{+20}$ MeV and $120(10)$
MeV, on our lattices with $a$ = 0.09 and 0.06 fm respectively. 
We are also studying heavy-light decay
constants. Necessary renormalization constants will be calculated in
future.  Our preliminary results for $f_{D^*_s}/ f_{D_s}$ (for which
the renormalization constants approximately cancel) is 1.15(10), from
our coarser lattice.
 \vspace{-0.15in}
 \section{Acknowledgement}
The computations were carried on the Blue Gene P of Indian Lattice Gauge Theory
Initiative, Tata Institute of Fundamental Research (TIFR), 
Mumbai, and on the Konark cluster, NISER. We would like to thank A. Salve and K. Ghadiali for technical support. We are
grateful to the MILC collaboration and in particular to S. Gottlieb, for
providing us with the HISQ lattices. We also thank A. Li and A.
Alexandru for help with numerical issues of overlap fermions.


\vspace{-0.15in}
\begin{thebibliography}{99}
\bibitem{HISQ_MILC} A. Bazavov {\it et al.}  Phys. Rev. D 82, 074501 (2010); A. Bazavov {\it et al.}, PoS(Lattice 2010)320 (2010); A. Bazavov et al. (MILC Collaboration), PoS(Lattice2012)158.
\bibitem{HISQ_action} E. Follana {\it et al.}, Nucl. Phys. B (Proc. Suppl.) 129 and 130, 447 (2004); E. Follana {\it et al.}, Phys. Rev. D75, 054502 (2007).
\bibitem{overlap} H. Neuberger, Phys. Lett. B417 (1998) 141; {\it ibid.} B427 (19998) 353.
\bibitem{overlap_chsymmetry} M. Luscher, Phys. Lett. B428 (1998) 342.
\bibitem{overlap_multimass} R.	Edwards {\it et al.},  Phys.Rev. D59 (1999) 094510.
\bibitem{overlap_dynamical} H. Fukaya {\it et al.}, [JLQCD Collaboration], Phys. Rev. Lett. 98, 172001 (2007); Phys. Rev. D 77, 074503 (2008).
\bibitem{overlap_chiQCD1} A. Li {\it et al.} Phys. Rev. D82 (2010) 114501;  N. Mathur {\it et al.}, PoS LATTICE2010 (2010) 114.
\bibitem{overlap_chiQCD} Y. Chen {\it et al.}, Phys. Rev. D70 (2004) 034502, S.J. Dong et. al, Phys. Rev. Lett. 85 (2000) 5051-5054, {\it ibid.}  Phys.Rev. D65 (2002) 054507.
\bibitem{strange_tune} C.T.H. Davies {\it et al.}, Phys.Rev. D81 (2010) 034506.
\bibitem{overlap_quenched_disp} S. Tamhankar {\it et al.}, Phys. Lett. B638 (2006).
\bibitem{MILC_hfs} T. Burch {\it et al.}, Phys.Rev. D81 (2010) 034508; G.C. Donald {\it et al.}, Phys.Rev. D86 (2012) 094501.
\bibitem{davies_Ds_eta_c}C. Davies, {\it et al.}, Phys.Rev. D82, 114504 (2010).
\bibitem{charm_baryon_hwlin} N. Mathur {\it et al.}, Phys.Rev. D66 (2002) 014502; R. Briceno {\it et al.}, Phys.Rev. D86 (2012) 094504.
\bibitem{MAPQchpt} O. Bar {\it et al.}, Phys. Rev. D 67, 114505 (2003), ibid. 70, 034508 (2004); J.W. Chen {\it et al.},  Phys. Rev. D75. 054501 (2007); K. Orginos {\it et al.}, Phys. Rev. D77, 094505 (2008) and references therein.
\bibitem{delta_mix} M. Lujan {\it et al.}, Phys.Rev. D86 (2012) 014501.
\end{thebibliography}
\end{document}